\documentclass[journal]{IEEEtran}
\usepackage{epsfig,latexsym,amssymb,amsmath,graphics,color}
\usepackage{graphicx,subfigure}
\newtheorem{prop}{Proposition}

\newtheorem{cor}{Corollary}

\newtheorem{lm}{Lemma}

\newtheorem{thm}{Theorem}

\newcommand{\be}{\begin{eqnarray}}
\newcommand{\ee}{\end{eqnarray}}
\newcommand{\benn}{\begin{eqnarray*}}
\newcommand{\eenn}{\end{eqnarray*}}
\def\IR{\rm I \kern-0.20em R}

\newcommand{\bthm}{\begin{thm}}
\newcommand{\ethm}{\end{thm}}

\newcommand{\bcor}{\begin{cor}}
\newcommand{\ecor}{\end{cor}}
\newcommand{\bprop}{\begin{prop}}
\newcommand{\eprop}{\end{prop}}
\newcommand{\blm}{\begin{lm}}
\newcommand{\elm}{\end{lm}}
\newcommand{\beq}{\begin{equation}}
\newcommand{\eeq}{\end{equation}}
\newcommand{\ber}{\begin{eqnarray}}
\newcommand{\eer}{\end{eqnarray}}

\newcommand{\bproof}{\begin{proof}}
\newcommand{\eproof}{\end{proof}}

\newcommand{\bit}{\begin{itemize}}
\newcommand{\eit}{\end{itemize}}
\newcommand{\ben}{\begin{enumerate}}
\newcommand{\een}{\end{enumerate}}
\newcommand{\bdesc}{\begin{description}}
\newcommand{\edesc}{\end{description}}
\newcommand{\beqarrn}{\begin{eqnarray*}}
\newcommand{\eeqarrn}{\end{eqnarray*}}
\newcommand{\bproofof}{\begin{proofof}}
\newcommand{\eproofof}{\end{proofof}}
\newenvironment{rem}{\begin{trivlist}\item[]{\bf
Remark:}\hspace{4mm}}{\end{trivlist}}
\newcommand{\brem}{\begin{rem}}
\newcommand{\erem}{\end{rem}}
\newenvironment{rems}{\begin{trivlist}\item[]{\bf
Remarks}\begin{itemize}}{\end{itemize}\end{trivlist}}
\newcommand{\brems}{\begin{rems}}
\newcommand{\erems}{\end{rems}}
\newtheorem{fact}{Fact}
\newcommand{\bfact}{\begin{fact}}
\newcommand{\efact}{\end{fact}}
\newtheorem{examp}{Example}
\newcommand{\bexamp}{\begin{examp}\rm}
\newcommand{\eexamp}{\end{examp}}
\newtheorem{defn}{Definition}
\newcommand{\bdefn}{\begin{defn}\rm}
\newcommand{\edefn}{\end{defn}}

\newtheorem{alg}{Algorithm}
\newcommand{\balg}{\begin{alg}}
\newcommand{\ealg}{\end{alg}}

\newtheorem{prob}{Problem}
\newcommand{\bprob}{\begin{prob}}
\newcommand{\eprob}{\end{prob}}

\newcommand{\bvtm}{\begin{verbatim}}
\newcommand{\bfig}{\begin{figure}}
\newcommand{\efig}{\end{figure}}
\newcommand{\bcen}{\begin{center}}
\newcommand{\ecen}{\end{center}}

\long\def\comment#1{}

\def \n2{{N_0 \over 2}}

\def \h5{\hspace{0.5in}}

\begin{document}

\title{Experimental MIMO VLC Systems Using Tricolor LED Transmitters and Receivers}
\author{
Shangbin Li, Boyang Huang, and Zhengyuan Xu
\thanks{This work was supported by National Key Basic Research Program of China (Grant No. 2013CB329201), Key Program of National Natural Science Foundation of China (Grant No. 61631018), National Natural Science Foundation of China (Grant No. 61501420), Key Research Program of Frontier Sciences of CAS (Grant No. QYZDY-SSW-JSC003), Key Project in Science and Technology of Guangdong Province (Grant No. 2014B010119001), and Shenzhen Peacock Plan (No. 1108170036003286).
}
\thanks{S. Li, B. Huang and Z. Xu are with Key Laboratory of Wireless-Optical Communications, Chinese Academy of Sciences, University of Science and Technology of China, Hefei, Anhui 230027, China. Z. Xu is also with Shenzhen Graduate School, Tsinghua University, Shenzhen 518055, China. Email: \{shbli, xuzy\}@ustc.edu.cn.}}

\maketitle
\thispagestyle{empty}
\pagestyle{empty}

\begin{abstract}
This paper shows experimental systems using R/G/B LEDs as both the emitters and narrowband filterless photodetectors to realize multiple-input multiple-output (MIMO) visible light communications. Among $9$ tested color-pair links from R/G/B LEDs to R/G/B LEDs, the R-R link and B-G link exhibit the best quality. With a $2\times2$ MIMO setup, the implemented OOK system can achieve a data rate of $40$kbps and $20$ kbps under half duplex and full duplex designs respectively, without any amplifying and equalizing circuits. When LEDs are used as photodetectors, the measured LED impedance spectrum versus frequency implies that the performance of the LED-to-LED communication link can be significantly improved via an impedance-matching amplifying and equalization circuit. Capabilities of simultaneous transmission and reception by an LED without electronic switching are experimentally observed. This feature enables continuous full duplex communication and offers an advantage over time division multiplexing.
\end{abstract}
\renewcommand{\IEEEkeywordsname}{Index Terms}
\begin{IEEEkeywords}
LED receiver, full duplex, VLC, multiple-input multiple-output.
\end{IEEEkeywords}
\IEEEpeerreviewmaketitle
\section{Introduction}

The light-emitting diodes (LEDs) are typically used as emitters in visible light communication (VLC), while the silicon-based PIN photodiodes (PIN-PDs) or the avalanche photodiodes (APDs) are adopted as the receiver \cite{1,2,3}. In order to improve the data rate of the white LED-based VLC link, a blue-pass filter is usually attached to the receiver in order to obtain high signal to interference plus noise ratio (SINR).

In addition to electrical-to-optical (EO) conversion, the LEDs can also realize optical-to-electrical (OE) conversion. This means they can serve as optical signal detectors as well, like the Si-PD. The time-division half duplex LED-to-LED VLC has been demonstrated \cite{4,5,6,7}. So far, only the red and amber LEDs have been considered in the high speed VLC receiver. Many other LEDs belong to the wide-bandgap optoelectronic devices, such as InGaN/GaN blue and green LEDs and the AlInGaP red LEDs. Both the emitting wavelength and photoelectric responsive wavelength decrease with the width of bandgap. The photons with energy greater than the bandgap can excite the electron-hole pairs. These pairs then move along the opposite directions in the active layer of the LEDs.

The commercial blue/green LEDs typically consist of n-GaN, InGaN multi-quantum-well (MQW), and p-GaN layers. In addition, there is a GaAlN blocking layer between the InGaN MQW and the p-GaN layer to avoid the electron leakage. The InGaN MQW structures in LEDs modify the density distributions of both the electrons and holes. It is conjectured they could affect the optical response properties if compared with the InGaN photodiode without the MQW. The MQW structure confines the electrons and holes, and weakens their drift.

Our preliminary experimental results show the luminous efficacy and photoelectric responsivity of the LEDs may be mutually offensive, i.e., it is difficult to simultaneously improve the EO and OE conversion efficiencies of the LED. Due to the orders of magnitude difference between the mobilities of the electrons and the holes, the OE conversion efficiency is much smaller than the EO conversion efficiency. There are similar properties for the AlInGaP red LEDs. Although the commercial high brightness InGaN/GaN or AlInGaP LEDs do not simultaneously combine the high external EO and OE conversion quantum yields, the LED with the InGaN/GaN multiple quantum dot structure shows the maximum responsivity of $0.13A/W$ for its photodiode characteristics in reverse bias \cite{8}.

Here, we prepare a $3\times3$ R/G/B LED module consisting of $3$ LEDs in each color. LEDs with the same color are connected in series, and different color LEDs are electrically independent. Pair-wise responses of different color LEDs are measured. Besides the filterless band response properties, the LED array exhibits a unique phenomenon different from the Si-PD array. The signal distortion in the RGB LED array due to partial obstruction of the link heavily depends on the number and color of the underlined LEDs. The experimental results indicate that the $3\times3$ RGB LED module can be utilized as the spectrum-tunable and narrow-band receiver. The spectrum tunability can be realized via adjusting the reverse bias of the individual channel in the module. Such two RGB LED modules form a pair of transceivers to set up a low complexity full-duplex or a $2\times2$ MIMO half-duplex LED-LED VLC system without external optical filters. The implemented system demonstrated a data rate up to $40$kbps.

The measured system performance can meet the narrowband application requirement, such as narrowband internet of things (NB-IoT), via adopting the simplest on-off keying (OOK) modulation. It does not require any equalizing and amplifying circuit. The performance can be significantly improved if an impedance-matching amplifying circuit is adopted.

The LED-LED visible light communication has the potential application in NB-IoT, and to some extent, can provide better downlink and uplink reciprocity than the LED-PD link. The previous researches about the duplex visible light communications based on the LED-LED link usually adopt the time division duplex (TDD). However, we experimentally demonstrate the simultaneous full duplex capability of the LED, i.e., the LED can receive the modulated optical signal when it is lighting for signal transmission. No time division is necessary.

\section{Experimental RGB LED receiver responses to visible light}

As receivers, the LED responses to visible light are experimentally measured. First we use the Luxeon Rebel red (LXML-PD01), green (LXML-PM01) and blue (LXML-PB01) LEDs as both the light source and light receiver. As shown in Fig. \ref{LED}, the LED module has a $3\times3$ R/G/B square array with $30$mm of the lattice spacing, where three same-color LEDs are connected in series as an array. Each LED is covered by a lens with the full beam angle of $12^\circ$. Ordinarily, the lens can provide significant channel gain of the LED-LED link. The Luxeon Rebel series green and blue LEDs adopt the flip-chip package of the InGaN dice in parallel connection with a transient voltage suppressor (TVS). The Luxeon Rebel series red LEDs adopt the chip-on-ceramic-board package of the AlInGaP dice.

\begin{figure}[h]
\centering
\includegraphics[width=0.4\columnwidth]{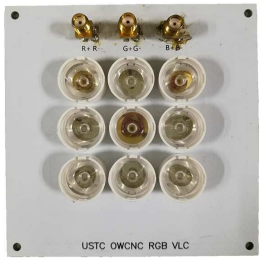}
\caption{Picture of the $3\times3$ R/G/B LED module.}
\label{LED}
\end{figure}

Two LED modules in parallel are pointing to each other, creating 9 possible color-pairs since each side has red, green or blue arrays. The output signals from LEDs at the receiver side are recorded by an oscilloscope (Keysight MSO-X 6004A). Totally 9 groups of link experiments are conducted: R-R, R-G, R-B, G-R, G-G, G-B, B-R, B-G, B-B.
The distance between the transmitter (Tx) and receiver (Rx) is around $10$cm. Any potential electromagnetic interference (EMI) between the Tx and the Rx has been carefully dealt with. And all the LEDs operate at a normal voltage with peak-to-peak voltage 1V. The measurement load of the oscilloscope is conveniently set as $1M\Omega$ for amplifying the OE conversion current signals of the LEDs. We do not utilize any impedance-matching amplifying circuit in order to avoid the external influence outside the LEDs as far as possible.

\subsection{RGB LED response to RGB LED}
The photodiodes or solar cell based on the AlInGaP or InGaN/GaN have attracted much attention during the past decade. The photoluminescence (PL) with photo absorption and electroluminescence (EL) through current injection of both the AlInGaP and InGaN/GaN LEDs have also been comparatively investigated \cite{9,10}. Being different from the previous results, our experimental results show that red LED can respond to red and green light, and has no response to blue light. It is conjectured the insensitivity of the Luxeon Rebel red LED to the blue light results from the packaging structure, in which the cathode of the red LED is upward and most of the injected blue light is absorbed by the n-GaP and can not effectively excite the electron-hole pair in the active layer. Figure \ref{rgb2rgb}(a) presents the results of red LED response to red LED with sine wave of frequency 10kHz. Compared with the results of red to red, the red LED response to green LED is much smaller and can be ignored.
When using green LED as the receiver, it has no response to red light. Moreover, unlike the red LED, green LED has little response to itself, but has strong response to blue LED as shown in Fig. \ref{rgb2rgb}(b).

In some aspects, blue LED is regarded as the best receiver since blue LED is only responsive to blue light. However, its response strength is not so good as compared with green LED responding to blue light.

\begin{figure}[!ht]
\centering
\subfigure[]{\includegraphics[width=0.45\columnwidth]{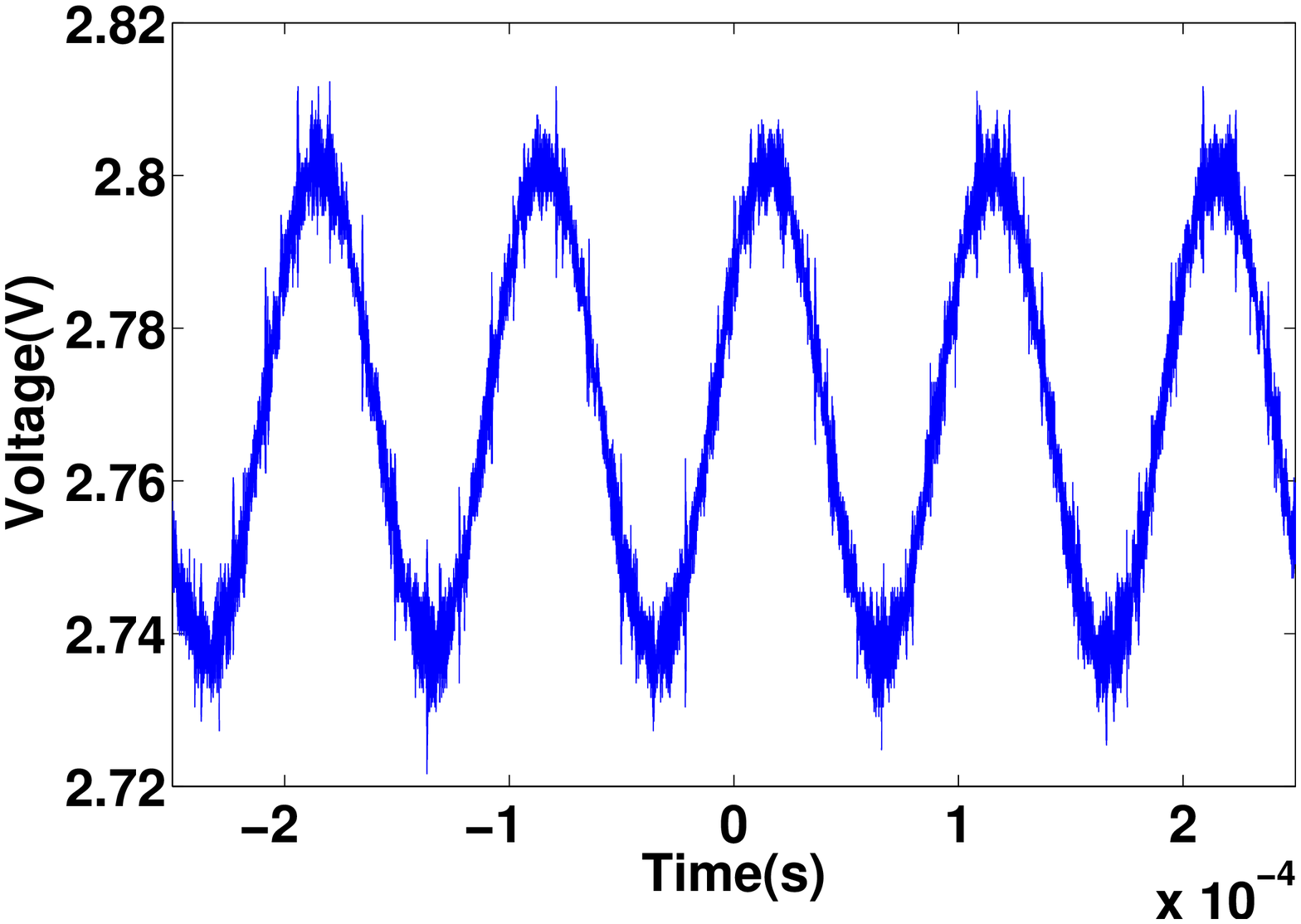}}
\subfigure[]{\includegraphics[width=0.45\columnwidth]{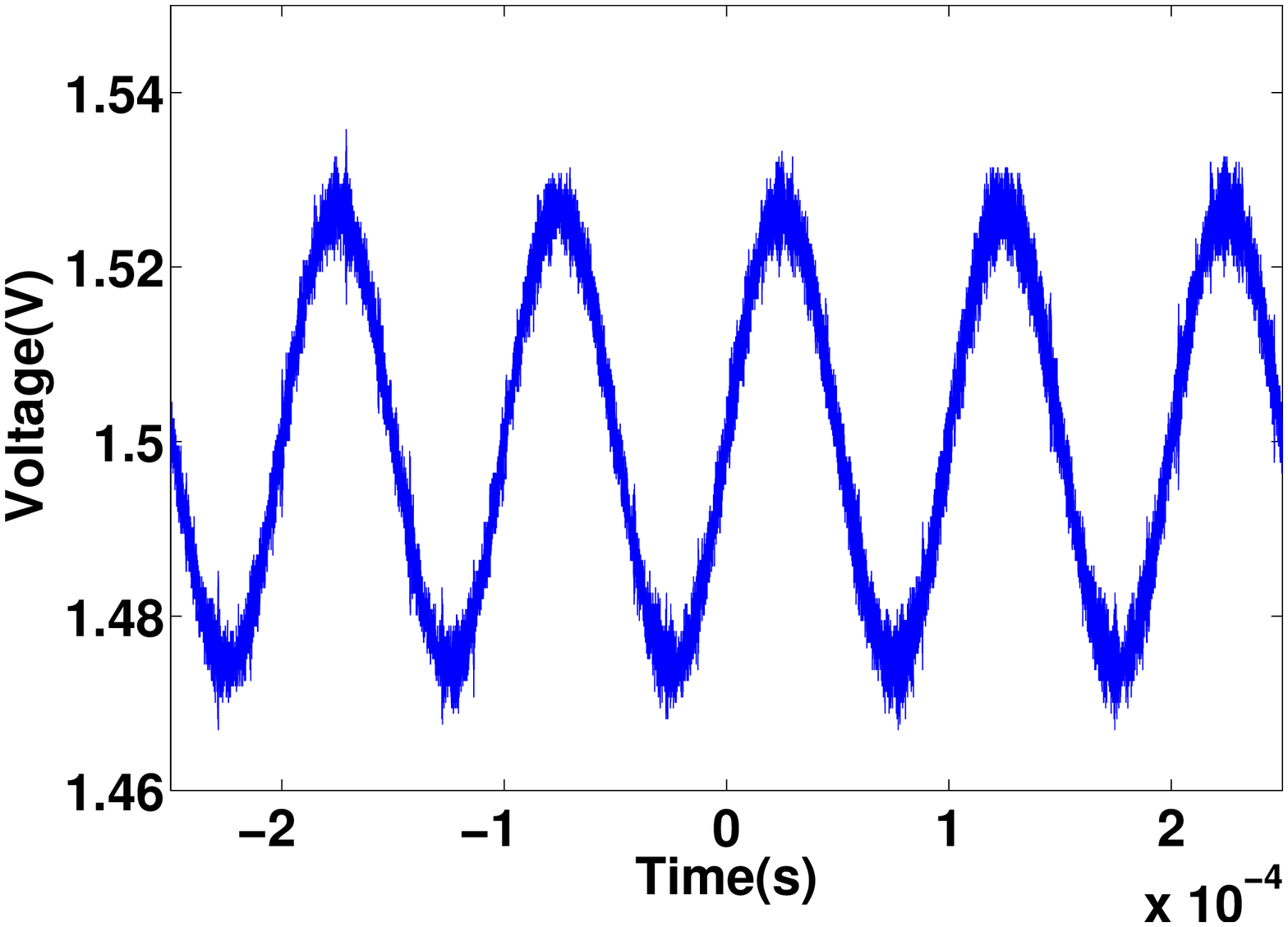}}
\caption{Some response of LED-to-LED with 10kHz sine wave. (a) Red to red LED response, (b) green to blue LED  response.}
\label{rgb2rgb}
\end{figure}

\subsection{The influence of partial obstruction on the photo response of the LED array}

When the $3\times3$ RGB LED module is used as the photodetector, it can be regarded as certain kind of photodetector array. However, if compared with the Si-PD array, the LED array exhibits some unique phenomena about the photo response due to its wide direct bandgap. We have observed the influence of the partial obstruction on the signal received by the $3\times3$ RGB LED module, and found the following interesting points:

\begin{itemize}
\item If one or two arbitrary red LEDs acting as optical receiver are blocked, both the received DC and AC current signals disappear;

\item if one arbitrary green LED acting as optical receiver is blocked, the received DC signal decreases to zero but AC current signal is robust. If two arbitrary green LEDs are blocked, no signal appears;

\item if one arbitrary blue LED acting as optical receiver is blocked, the received DC signal is cut to the half value but AC current signal is robust. If two arbitrary blue LEDs are blocked, the DC signal disappears and the AC signal decreases over the half value.
\end{itemize}

It is analyzed that the above phenomenon results from two facts: one is whether the LED is packaged with the TVS or not, and the other is the match between the bandgap of the LED and the threshold voltage of the TVS. This property of the RGB LED array is very valuable in the color interference management of the VLC communication system.

\subsection{RGB LED responses to various colors}
Marcin Kowalczyk mentioned \cite{7} that an LED can detect light whose wavelength is shorter than its emission wavelength within 100nm. The LED structure  is similar to a photodiode (PD), as both are a PN junction. Two effects constrain the responsive wavelength. The bandgap determines the LED emission wavelength as the transmitter, as well as upper cut-off wavelength as the receiver. Any photon wavelength larger than that does not have enough energy to excite electron-hole pairs. And the shorter wavelength is mainly limited by material absorbtion. The absorbtion coefficient increases with decreased wavelength \cite{11}. When wavelength is short, photons are more likely absorbed by the PN junction surface and make little contribution to the photocurrent. These two effects make LED a narrowband photon detector and even narrower than most commercial photodiodes.

\begin{figure}[!ht]
\centering
\subfigure[]{\includegraphics[width=1.7in]{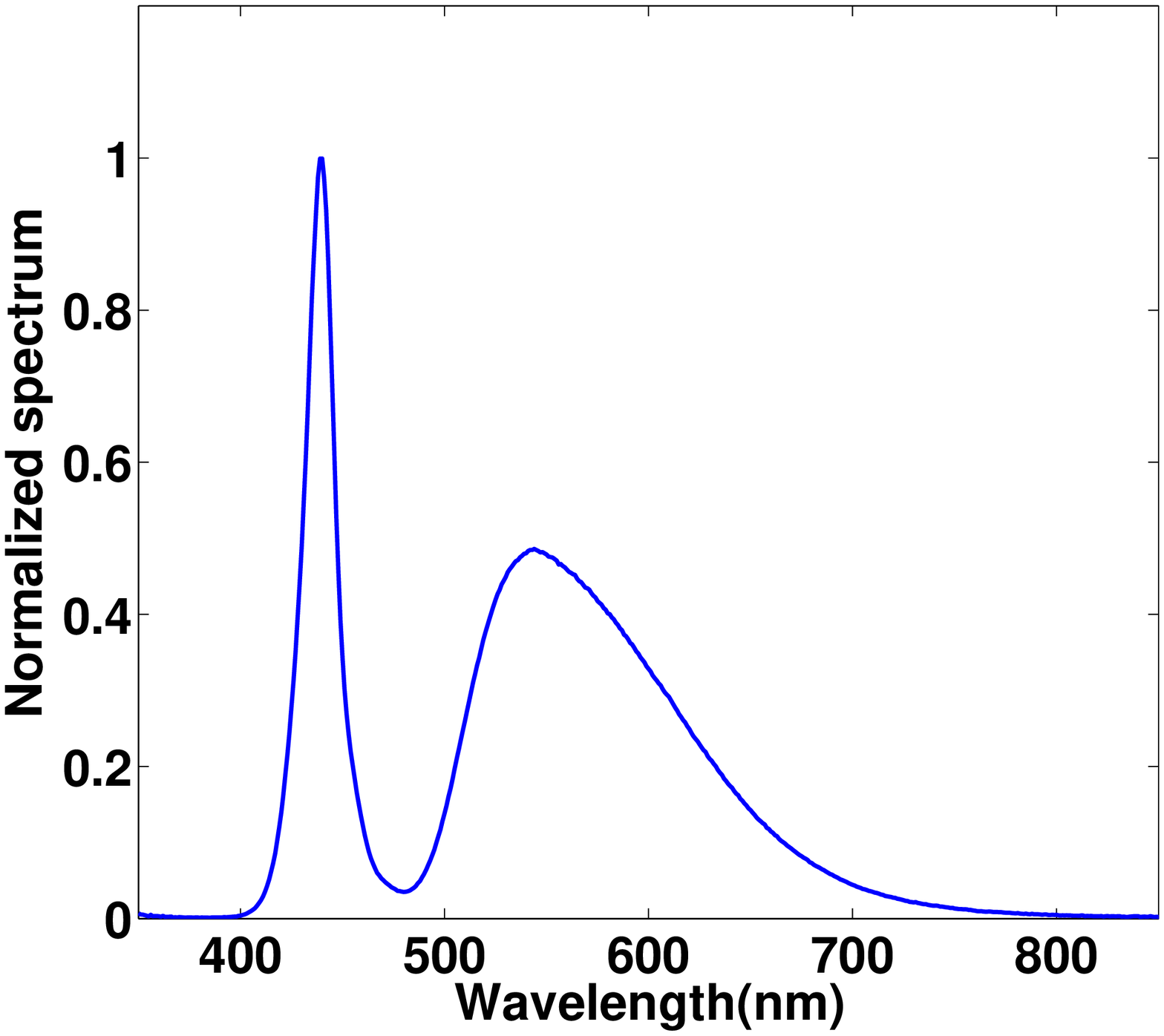}}
\subfigure[]{\includegraphics[width=1.7in]{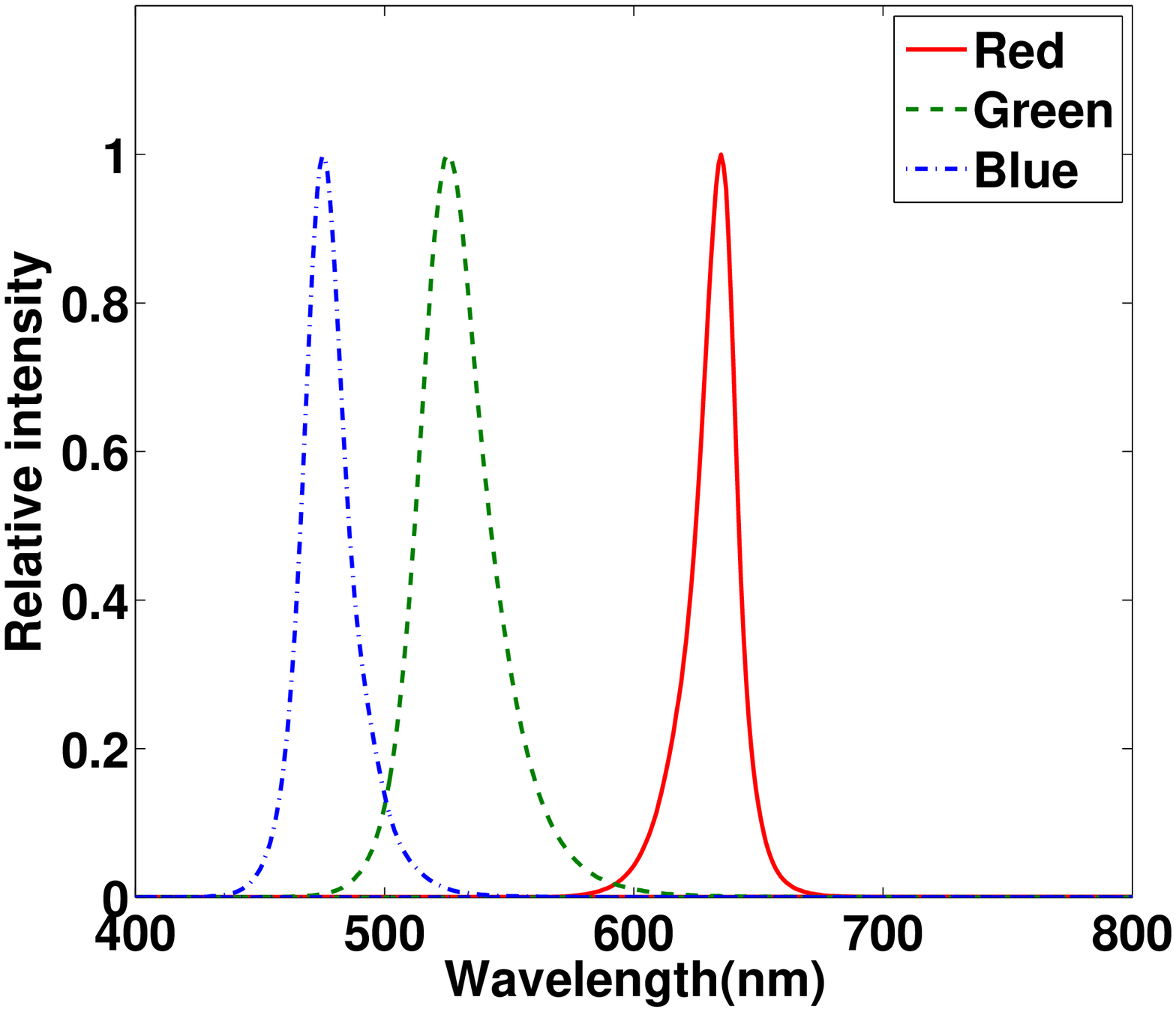}}
\caption{Spectrum of emitting LEDs. (a) White LED. (b) R/G/B LEDs with peak wavelength and FWHM pairs: (635, 17.3)nm, (525, 30.4)nm, (475, 19.5)nm.}
\label{spectrum}
\end{figure}

We test a wide range of spectral responses of the RGB LEDs to different input colors, using a white light source and an optical filter over the receiving LED to select proper wavelength band of the light impinged upon the LEDs. White light spectrum ranges from 400nm to 800nm as seen in Fig. \ref{spectrum}(a). According to emission spectrum of RGB LEDs in Fig. \ref{spectrum}(b), filters are chosen as follows (center wavelength/FWHM in nm): 525/50, 582/75, 630/38 for red LED; 435/40, 475/50, 525/50 for green LED; 435/40, 475/50 for blue LED. Transmitting white light LED is driven by a constant voltage, and the response of receiving LED is measured in terms of voltage by the oscilloscope. Each response of an LED with a filter is divided by the optical power impinged upon the LED. This ratio is further normalized by the ratio without a filter (in this case, the LED response to the whole white spectrum is measured), yielding the relative response ratio. Thus the relative response ratio without a filter is defined as 1. When an LED's relative response ratio is larger than 1, it means the average contribution of impinging light within the corresponding filter's transmission range is larger than the average contribution of the whole spectrum. And the larger the ratio, the more sensitive of the LED in this band.

Tables \ref{tab1}-\ref{tab3} present the results for R/G/B LED receivers respectively. The red LED has strong responses to light bands 582/75nm and 630/38nm, and the blue and green LEDs show strong responses to light bands  435/40nm and 475/50nm. Compared with green LED, the OE response wavelength of blue LED tends to be shorter. Always using same color LEDs as both the transmitter and receiver may not be optimal, because its EO emission and OE response bands are usually mismatched. However, mismatch is small for some colors, such as red LED. Also, one can find a suitable pair of colors for a transmitter-receiver pair, for example, blue LEDs emission and green LEDs detect.

\begin{table}[!ht]
\centering
\caption{Red LED receiver with filters}
\label{Red_filter}
\begin{tabular}[t]{|c|c|c|c|c|}
\hline
Filter&None&525/50&582/75&630/38\\
\hline
Response voltage(V)&4.56&0.50&3.61&1.24\\
\hline
Relative response ratio&1&0.46&2.19&2.64\\
\hline
\end{tabular}\label{tab1}
\end{table}

\begin{table}[!ht]
\centering
\caption{Green LED receiver with filters}
\label{Green_filter}
\begin{tabular}[t]{|c|c|c|c|c|}
\hline
Filter&None&435/40&475/50&525/50\\
\hline
Response voltage(V)&3.03&2.2&0.56&0.1\\
\hline
Relative response ratio&1&3.01&2.99&0.13\\
\hline
\end{tabular}\label{tab2}
\end{table}

\begin{table}[!ht]
\centering
\caption{Blue LED receiver with filters}
\label{Blue_filter}
\begin{tabular}[t]{|c|c|c|c|}
\hline
Filter&None&435/40&475/50\\
\hline
Response voltage(V)&4.7&3.3&0.61\\
\hline
Relative response ratio&1&2.89&2.16\\
\hline
\end{tabular}\label{tab3}
\end{table}

\section{Experimental performances of full duplex or MIMO VLC systems using LED transmitters and receivers}

From the previous experimental results, we find two possible R-R and B-G links nearly free of interference to each other. Thus, we implement a full-duplex system and a $2\times2$ MIMO system, using red and blue LEDs as transmitters, and red and green LEDs as receivers.  In the following experiment, we use $3\times3$ RGB LED arrays as both transmitters and receivers. Each color array has three LEDs connected in series and transmitting/receiving the same signals.

Figure \ref{duplex} shows the diagram of the full-duplex communication system. The forward link is from the red LED array to the red LED array, and reverse link from the blue LED array to the green LED array. Two LED arrays on the same link are separated by 10cm, and there is a convex lens over each LED to converge light. OOK signals are sent to the transmitting LED array through an RIGOL DG5252 arbitrary waveform generator (AWG) with bias voltage $V_{dc}=8V$ and peak-peak voltage $V_{pp}=1V$. The received signals output by the receiving LED array are recorded by Agilent MSO-X 6004A Oscilloscope and the bit error rate (BER) is calculated off-line. The MIMO LED-to-LED communication system is similar except one directional link from one color to the other color, as shown in Fig. \ref{MIMO}.
Figure \ref{BER} shows the measured BER performance for data rate from 10kbps up to 50kbps. The R-R link shows the minimum error which is determined by the number of transmitted bits (16384), and the B-G link gives higher BERs.

\begin{figure}[!ht]
\centering
\includegraphics[width=0.35\columnwidth,angle=-90]{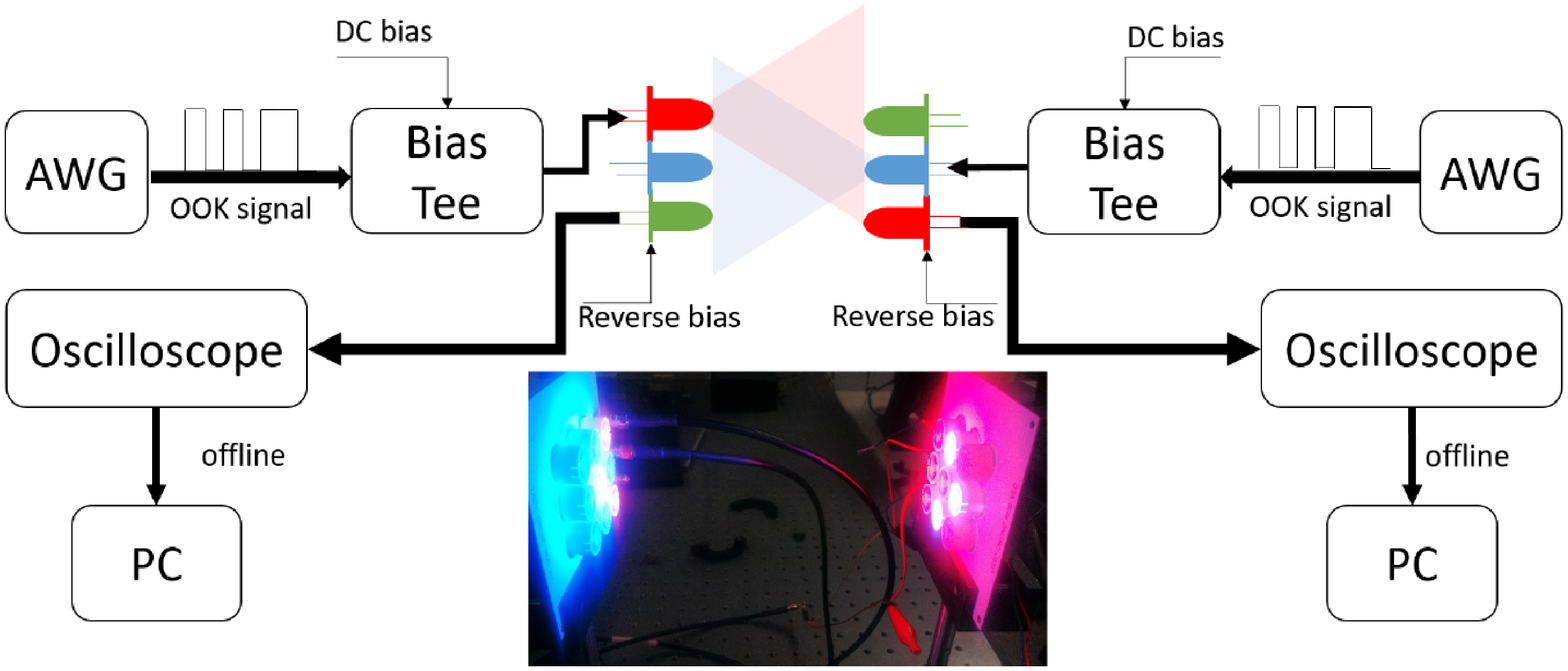}
\caption{Full-duplex LED-to-LED communication using two links: red LED array to red LED array and blue LED array to green LED array.}
\label{duplex}
\end{figure}

\begin{figure}[!ht]
\centering
\includegraphics[width=3.2in]{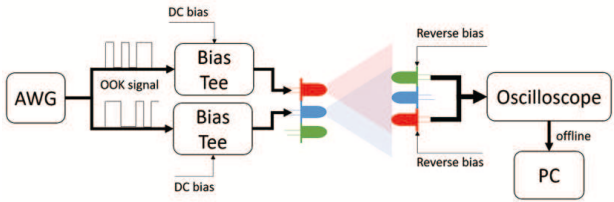}
\caption{$2\times2$ MIMO LED-to-LED communication using R/B LEDs as transmitters and R/G LEDs as receivers.}
\label{MIMO}
\end{figure}

\begin{figure}[!ht]
\centering
\includegraphics[width=2.8in]{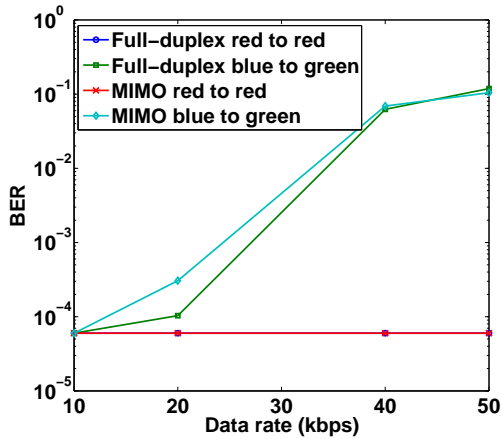}
\caption{Measured BERs of full-duplex and MIMO systems versus data rate.}
\label{BER}
\end{figure}

The impedance mismatch may affect the received signal strength. The input impedance of the oscilloscope was set as $1M\Omega$, to convert the receiving LED output current to a large voltage as compared with impedance setting of $50\Omega$. Figure \ref{impedance} provides the impedance magnitude of red and green LEDs, measured by a network analyzer. The real part of the impedance is resistance. The green LED has smaller resistance than red LED, and is perhaps more sensitive to impedance mismatch. Furthermore, Fig. \ref{3dBBW} gives the changes of 3dB bandwidth of the LED-LED link with the different resistance at the LED receiver. The 3dB bandwidth was obtained from the voltage signal across the load resistor which is connected in series with the LED receiver. It can be seen that the impedance indeed has influence on the data rate.


\begin{figure}[t]
\centering
\includegraphics[width=2.7in]{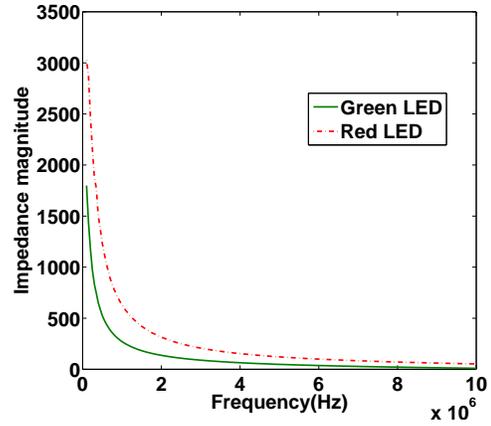}
\caption{Impedance of red and green LEDs as receivers with different frequency.}
\label{impedance}
\end{figure}

\begin{figure}[t]
\centering
\includegraphics[width=2.7in]{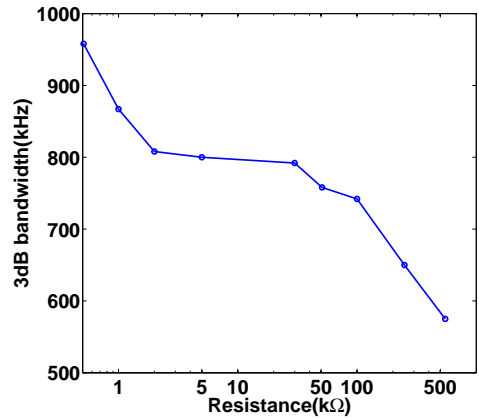}
\caption{The 3dB bandwidth of the LED-LED link with different series resistance.}
\label{3dBBW}
\end{figure}

\section{Simultaneous full duplexing for the LED-LED link}

The previous results demonstrate that LEDs can be used as VLC receivers. Thus a duplex link can be built using LEDs only. One traditional way to realize full duplexing is to use TDD. That is, the LEDs are forward or reverse biased respectively in different time slots on a time-sharing basis. However, in a realistic LED lighting application, LED light  is anticipated to be always on without any interruption. Hence, an LED is desirable to simultaneously receive signal while emitting light signal.

Here, we present a preliminary experimental demonstration that the LED can still respond to the optical signal when the LED is light-emitting. Our experimental setup uses a red LED simultaneously as the transmitter and receiver, and uses a bias-T device to separate the DC voltage from the AC response signal. The amplitude of the response signal is recorded by an oscilloscope. We experimentally measured the response signal when the LED forward voltage ranges from $1.4V$ to $1.74V$, representing a changing transmitted signal. The amplitude of the response signal is normalized over the response amplitude of no forward bias voltage. As can be seen from Fig. \ref{Resp-bias},  when the LED forward voltage is greater than $1.64V$, the response signal intensity begins to exponentially decay, fitting by a curve $e^{-26.5V_{dc}+44.4}$ for the voltage range of interest.

The relationship between the forward driving current of the red LED and the response signal intensity is also obtained, and shown in Fig. \ref{Resp-current}. By comparing the data, we find that the reciprocal of the response amplitude is linear with the forward current when the current is small. The above experimental results show the LED-LED simultaneously full duplex communication is possible. As the LED acts as the transmitter, the transmitted signal is known. Even if the sending signal and response signal interfere each other, the interference signal can be easily eliminated. But if the forward voltage of the LED is large, the amplitude of the response signal is very small which results in a low signal to noise ratio (SNR). Conversely, if the forward voltage of the LED is small, the SNR will not be very high due to the small amplitude of the signal. Thus, it is necessary to choose a suitable voltage to maximize the SNR. This preliminary observation opens a possibility for further investigation of simultaneous full duplex communication.

\begin{figure}[t]
\centering
\includegraphics[width=2.7in]{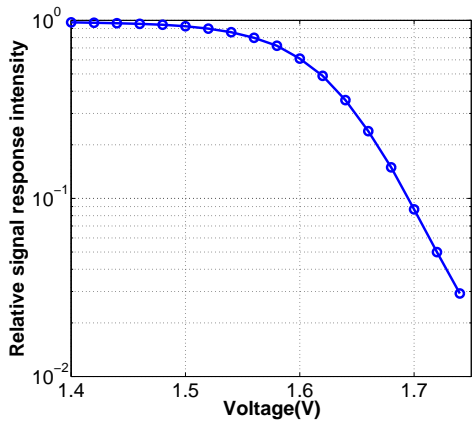}
\caption{The normalized response intensity versus the forward voltage.}
\label{Resp-bias}
\end{figure}

\begin{figure}[t]
\centering
\includegraphics[width=2.7in]{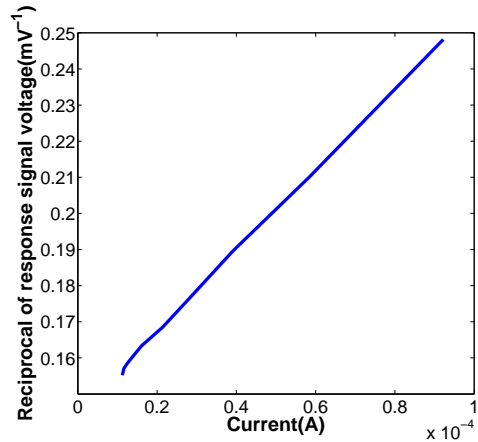}
\caption{The reciprocal of the injected optical signal response voltage of the red LED versus its forward current.}
\label{Resp-current}
\end{figure}

\section{Conclusion}
This paper has presented experimental results using R/G/B LEDs as both the transmitters and receivers. LED responses to different colors are reported. It is found that LED is a narrowband receiver. The red-red and blue-green links have no mutual interference even without an optical filter. An experimental $2\times2$ MIMO VLC system is further proposed, and a minimum of data rate 20kbps using OOK modulation is achieved at the BER below $10^{-3}$. Finally, we find that a full-duplex VLC system without time sharing is possible based on our preliminary investigation.

In the future,  the LED receiver system will be optimized jointly with the LED transmitter pre-distortion and receiver post-equalization in order to increase the data rate. A full duplex communication system allowing transmitter LED forward voltage to carry communication signals will be built and tested.

\bibliographystyle{IEEEtran}

\end{document}